# Highly-directional, highly-efficient solution-processed light-emitting diodes of all-face-down oriented colloidal quantum wells

*Hamed Dehghanpour Baruj[†], Iklim Yurdakul[†], Betul Canimkurbey[†¥], Ahmet Tarik Isik[†], Farzan Shabani[†], Savas Delikanli[†§], Sushant Shendre[§], Onur Erdem[†], Furkan Isik[†], and Hilmi Volkan Demir[†§]*

[†] Department of Electrical and Electronics Engineering, Department of Physics, UNAM − Institute of Materials Science and Nanotechnology, Bilkent University, Ankara 06800, Turkey

[§] LUMINOUS! Center of Excellence for Semiconductor Lighting and Displays, School of Electrical and Electronic Engineering, School of Physical and Mathematical Sciences, Nanyang Technological University, Singapore, 639798, Singapore

[¥] Central Research Laboratory, Amasya University, Amasya, 05100, Turkey

## Abstract

Semiconductor colloidal quantum wells (CQWs) make an exciting quasi-2D class of nanocrystals thanks to their unique properties including their highly anisotropic optical transition dipole moment (TDM). Thus, employing a film of CQWs with face-down orientation as an emissive layer (EML) in an electroluminescent device is expected to substantially boost photon outcoupling efficiency. Here, we show all-solution-processed colloidal quantum well light-emitting diodes (CQW-LEDs) using a single all-face-down oriented self-assembled monolayer (SAM) film of CQWs that enables a high level of in-plane (IP) TDMs of 92%. This SAM film significantly enhances the outcoupling efficiency from 22% (of standard randomly-oriented emitters) to 34% (of face-down oriented emitters) and charge injection efficiency. This SAM-CQW-LED architecture enables a record high level of external quantum efficiency of 18.1% for the solution-processed type of CQW-LEDs, putting their efficiency performance on par with the hybrid organic-inorganic evaporation-based CQW-LEDs and all other best solution-processed LEDs. In addition, this architecture provides a high maximum brightness of 19,800 cd/m$^2$ with a long operational lifetime of 247 h at 100 cd/m$^2$ along with saturated deep-red emission (651 nm). These findings indicate the effectiveness of oriented self-assembly of CQWs as electrically-driven emissive layers in improving outcoupling and external quantum efficiencies in the CQW-LEDs.

**Introduction**

Semiconductor colloidal quantum wells (CQWs), alternatively nicknamed nanoplatelets (NPLs), having a quasi-two-dimensional structure with pure thicknesses, have emerged as an important class of nanocrystals (NCs)[1]–[4]. Compared to their spherical counterparts, NPLs exhibit a narrow emission bandwidth due to their suppressed inhomogeneous broadening, extraordinary large absorption cross-section owing to their giant oscillator strength, and highly suppressed Auger recombination thanks to their non-confined lateral dimensions[3]–[16]. These features have led to the fabrication of high external quantum efficiency (EQE) and high-color purity colloidal quantum well light-emitting diodes (CQW-LEDs) using hybrid organic-inorganic devices structures. For example, in the previous work of our group, we reported a high EQE value of 19.2% with a maximum brightness of 23,490 cd/m$^2$ for the hybrid organic-inorganic inverted CQW-LEDs[17]. However, the efficiency of CQW-LEDs having only all-solution processed structures in their layers is still lagging far behind the state-of-the-art values obtained with the evaporation-based devices[15], [18],[19]. Since the solution-processable architecture is relatively easier to fabricate and cheaper than the evaporation-based ones, achieving a high EQE from all-solution-processed LEDs is of great interest. Thus far, Kelestemur *et al*. reported the highest EQE of 9.92% with a maximum brightness of 46,000 cd/m$^2$ for all-solution processed red-emitting CQW-LED[15]. However, this EQE value was insufficient to fulfill the industrial demands. The fundamental light-extraction efficiency limitation and exciton quenching of the zinc oxide (ZnO) layer impede the maximum EQEs to reach the state-of-the-art values in the case of all-solution processing[17], [20].

As a new strategy, we target to overcome the light extraction efficiency limitation by exploiting self-assembled monolayer (SAM) films of NPLs as an emissive layer (EML)[21]–[23]. In this regard, NPLs are highly advantageous owing to their anisotropic transition dipole moments (TDMs) [21]–[23]. The CQWs are from the family of anisotropic nano-emitters that exhibit almost all in-plane (IP) TDMs[5], [22]–[27]. Thus, an all-face-down ensemble film of CQWs can act as a highly anisotropic EML. The Fresnel equation suggests that, from the emitter point of view, the ray escape cone has an outcoupling angle of 30° to the surface normal[5], [22]–[24]. Accordingly, the density of light inside the cone determines the intensity of extracted light. It has been reported that the ratio of optical power inside the cone is higher in a device with directional EML than in the isotropic EML[5], [22], [23], [25]–[31].

To control the orientation of CQWs, many deposition methods such as inject printing[32], evaporation depositions[33], and self-assembly[23], [30], [34] have been introduced. Among them, the most feasible and practical technique that provides total control over a millimeter square or larger size film is the liquid-air interface self-assembly. The feasibility of this method on the deposition of SAM in all-face-down configuration across a large-area and uniform film out of colloidal quantum wells has been reported previously by our group[34]. Thus far, implementation of this deposition method in an active device has not been reported yet. Here, addressing the possible interaction of liquid-air interface self-assembly's sub-phase with device layers and creating a highly uniform single monolayer film of CQWs free from pinholes are the required criteria for achieving highly efficient devices fabricated using a single monolayer of NPLs.

In addition, self-assembly provides the lowest possible root-mean-square (RMS) roughness among the various deposition techniques yielding an RMS roughness comparable with the spherical nanocrystals[17], [35]. The film roughness, especially the EML film roughness, is a deterministic factor in the efficiency of the devices[13], [17], [36]–[38]. It is well known that the interlayer transfer of electrons occurs by the hopping process, which depends on the atomic-scale distance between layers and hence the roughness of EML[2], [39]. Thus, self-assembled NPLs as an EML provide unique advantages.

In this paper, we proposed and demonstrated a high-performance all-solution-processed CQW-LED having all-face-down SAM film of NPLs as an EML through precise control over the liquid-air self-assembly for the first time. Our optimized device architecture is made of consecutive layers of indium tin oxide (ITO, 100nm)/poly(ethylene dioxythiophene):polystyrene sulphonate (PEDOT:PSS, 30nm)/poly (N,N9-bis(4-butyl phenyl)-N,N9-bis(phenyl)-benzidine) (p-TPD, 20nm)/poly(9-vinyl carbazole) (PVK ,10nm)/NPLs (20nm)/$Zn_{0.95}Mg_{0.05}O$(30nm)/Al(100nm). Using back-focal plane (BFP) imaging techniques on our all-face-down SAM film of NPLs, we measured TDMs direction ($\Theta$) to be 92% in-plane (IP). Considering our device's structure, we calculated the outcoupling efficiency to be 34%, which is significantly higher than 22% of the conventional solution-processed structures with isotropic emitters as an EML and without any light extraction feature. Our SAM-CQW-LEDs reached an outstanding EQE of 18.1 along with a high luminance level of 19,800 cd/m$^2$ and a long lifetime of 247h at 100 cd/m$^2$ using ligand-exchanged $CdSe/Cd_{0.25}Zn_{0.75}S$ core/hot-injection shell (HIS) NPLs. These values are comparable to those of the best reported organic-inorganic hybrid CQW-LEDs, state-of-the-art OLEDs, PeLEDs, and

QLEDs. In this work, we have, therefore, overcome the fundamental outcoupling efficiency limitations commonly encountered in conventional colloidal device structures.

**Results and discussion**

In this study, we used high photoluminescence quantum yield (PLQY) core/hot-injection shell (HIS) NPLs previously developed by our group[12], [17], [40], consisting of square-shaped cores with a gradient shell and oleic acid (OA) and oleylamine (OLA) ligands. The shell of $Cd_{0.25}Zn_{0.75}S$ was grown on the CdSe cores by adding the cadmium and zinc precursors at room temperature and continuously injecting the sulfur precursor at higher temperatures (the hot-injection method). Details of the synthesis procedures of CQWs are given in the SI. The outer shell structure ensures the confinement of the excitons inside the core region to impede nonradiative recombination by surface traps[41]. The gradient shell structure also provides a low lattice mismatch between the core and the outer shell, which reduces interface defects and nonradiative recombination. Finally, the outmost layer consists of OA and OLA ligands[17], [38]. The role of these ligands is to allow for good solubility in the colloidal solution, which is a critical parameter for solution-processed fabrication techniques while passivating the surface defects. Thanks to their heterostructure, the obtained NPLs show a PLQY of 95% in the solution and 87% on the quartz substrate, which is the highest PLQY among all types of CQWs in the saturated deep-red color, with excellent chemical and optical stability [5], [6], [17], [40], [42],[43].

Our devices consist of indium tin oxide (ITO, 100nm)/poly(ethylene dioxythiophene):polystyrene sulphonate (PEDOT:PSS, 30nm)/poly (N,N9-bis(4-butyl phenyl)-N,N9-bis(phenyl)-benzidine) (p-TPD, 20nm)/ poly(9-vinyl carbazole) (PVK ,10nm)/NPLs (20nm) /$Zn_{0.95}Mg_{0.05}O$(30nm) /Al(100nm). (see figs. 1a-b). All the layers were deposited using the spin-casting technique in dynamic mode except for the PVK layer, which was on the static mode. Herein, we took advantage of both the high hole mobility provided by poly-TPD ($1\times10^{-4}$ cm$^{-1}$V$^{-1}$s$^{-1}$) and the deep highest-occupied-molecular-orbit (HOMO) energy level (-5.8 eV) provided by PVK. This bilayer structure accompanied with PEDOT:PSS provides a stepwise energy level alignment between ITO and EML (see fig. 1c). Moreover, as an electron transport layer (ETL), we used Mg-doped ZnO (5%) to diminish the exciton quenching[16], [44]–[46][47]. It is known that the direct contact of ZnO to NPLs quenches the emission due to the nonradiative energy transfer, but this phenomenon is significantly suppressed by doping Mg in the ZnO structure[44], [46]. To confirm this effect, we analyzed in-film PLQY of HIS NPLs on ZnO and Zn$_{0.95}$Mg$_{0.05}$O layers. We found that in-solution PLQY of 95% for the NPLs dropped to 55% on ZnO and 64% on Zn$_{0.95}$Mg$_{0.05}$O. Therefore, this quenching limits the maximum achievable EQE based on the following relation[17], [20].

$$EQE = \eta_{out} \cdot r \cdot q \cdot \gamma \tag{1}$$

Here, $\eta_{out}$ represents the extraction efficiency, $r$ denotes the fraction of excitons with the probability of radiative recombination, $q$ is the film PLQY, and $\gamma$ gives the charge injection efficiency inside the device. Since the *EQE* and $q$ are directly correlated, a decrease in $q$ reduces the EQE. Moreover, the charge balance can be justified by checking the electronic and hole-only devices' current density versus voltage (J-V) curves. The J-V curve of the electron-only device fabricated by Zn$_{0.95}$Mg$_{0.05}$O is closer to the hole-only device's J-V than the electron-only device manufactured by ZnO nanoparticles (see fig. S1). Also, the X-ray photoelectron spectroscopy (XPS) depth profile results (see fig. 1a) show some degree of oxidation in the Al and Zn$_{0.95}$Mg$_{0.05}$O interface. This is due to the aging process and could increase the efficiency of devices by creating an aluminum-oxide layer that acts as an electron blocking layer by balancing the charges[48], [49]. However, in our case, we have not observed such an effect. Here we conclude that utilizing the Zn$_{0.95}$Mg$_{0.05}$O nanocrystals as an ETL is preferable to the ZnO nanocrystals in our devices.

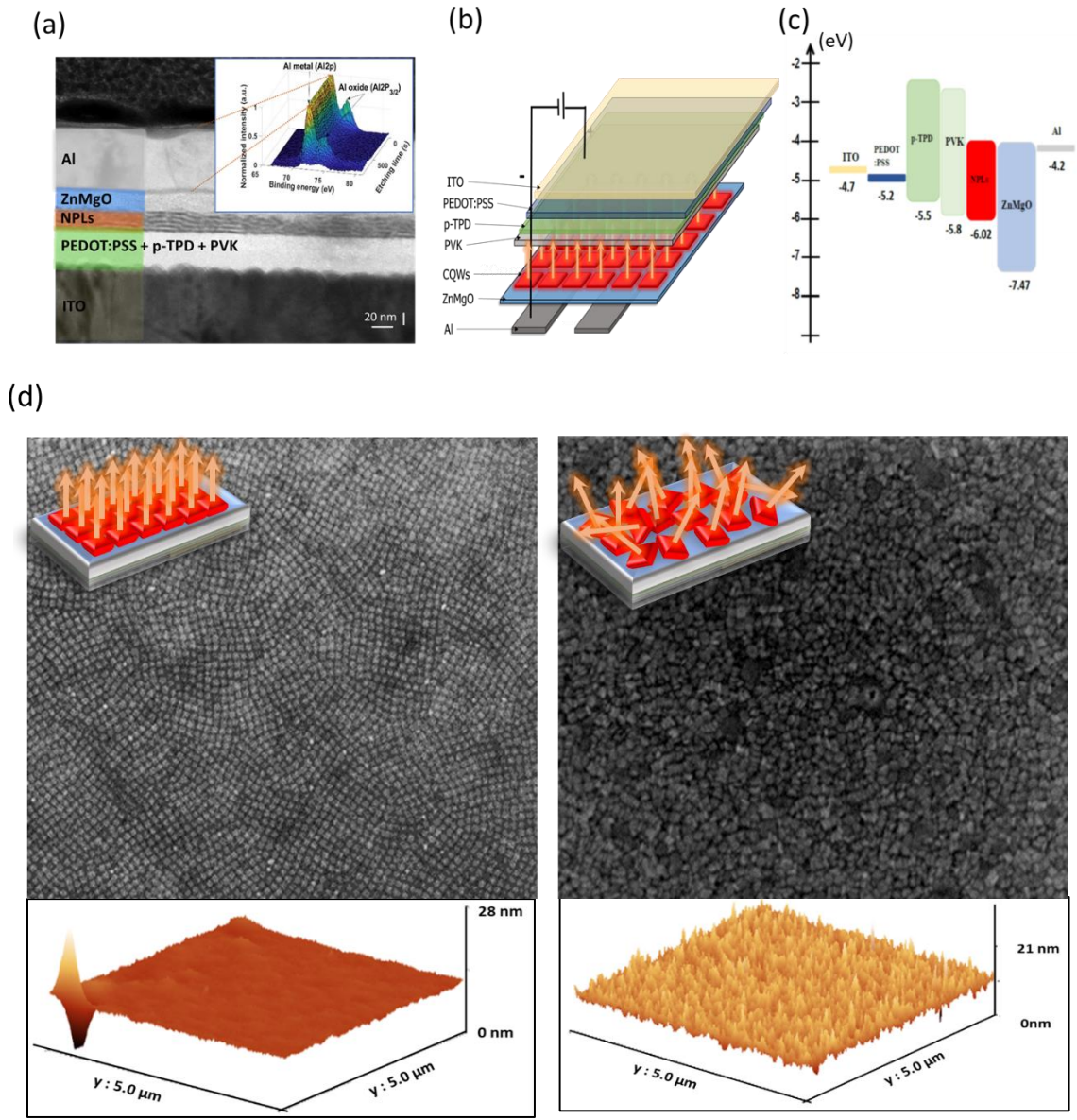

Fig. 1. a) Transmission electron microscopy (TEM) cross-sectional image of the device using the spin-coated hot-injection shell (HIS) NPLs; scale bar: 20 nm. Inset: 3D graph of X-ray photoelectron spectroscopy (XPS) depth profile of Al region with metallic and oxide Al peaks. b) Schematic of the device structure and c) energy band diagram of the optimized structure. d) Scanning electron microscopy (SEM) image of the all-face-down self-assembled monolayer with its atomic force microscopy (AFM) scan (left-hand side). Inset: the schematic of all-face-down SAM film of NPLs on the device's layers. SEM image of the spin-casted film of NPLs with the corresponding AFM scan (right-hand side). Inset: the schematic of a spin-casted film of NPLs on the device's layers.

To achieve a highly efficient device using a single monolayer assembly of oriented NPLs, the film should be highly uniform when all-face-down (see fig. 1d) but at the same time pinhole-free to avoid the charge leakage. Such a monolayer film can be achieved by precise control using the liquid-air self-assembly. In doing so, it is essential to prevent crack formation during the draining process, the pre-existed defects in the substrate, and the possible effects of subphases with the layers of the device. In this work, to be compatible with active device fabrication, we modified the liquid-air interface self-assembly approach our group previously developed to make passive film[34] and explored the applicability in the deposition of a highly uniform and pinhole-free all-face-down SAM film NPLs as an active film on top of the devices' layers. In this process, after coating the hole injection layer (HIL) and hole transport layers (HTLs), inside the $N_2$ filled glovebox, the substrates were moved into a Teflon container poured by the acetonitrile (ACN) subphase. Subsequently, 20 μL of NPL in hexane solution was dropped on the ACN from the edge of the container. They were dispersed across the surface of the subphase as soon as they fell on the ACN. After evaporation of hexane, a uniform all-face-down membrane of NPLs was obtained on top of the subphase. To transfer this film to the substrates, the ACN was drained through a needle from the bottom of the container, and the film sank onto the substrate coated with the PVK layer on the top. Note that for having a uniform film, the height of ACN on top of the substrate should be as low as possible to have the least amount of movement of the ensembled film while draining. Also, we have used silicon oil surfactant to compress the SAM membrane further and avoid the crack and void formation during the deposition process due to capillary forces by the walls of the container[34].

Fig. 1c shows the scanning electron microscopy (SEM) image of the mosaic-like structure of the SAM film, which is highly uniform with a complete surface coverage of NPLs on top of HTL compared to the spin-casted film that is randomly oriented. Inset schematics also illustrate the difference in the architecture of the films deposited by self-assembly (all-face-down oriented) and spin-casting (randomly oriented), respectively, with their corresponding emission polarization. Also, the atomic force microscopy (AFM) measurements reveal the RMS roughness <1 nm for the SAM films and 2.4 nm for the spin-coated ones (see fig. 1c). These results imply that the orientation-controlled film yields the lowest possible surface roughness, which is comparable with that of spherical nanocrystal film. This reduction in film roughness of NPLs should increase the charge injection.

In addition, to check the potential effects of ACN solvent on the electronic properties of HTLs and HIL, we fabricated hole-only devices with the structure of ITO/PEDOT:PSS/p-TPD/PVK/NPLs/Al. The layers were deposited by spin-coating process except for Al which was deposited by thermal evaporation. In this device, after depositing the ITO/PEDOT:PSS/p-TPD/PVK layers, we placed the sample inside the ACN for 5 min and then baked it for 30 min at 90 °C. Finally, the NPLs were coated on the samples, followed by Al deposition. As shown in Fig. S2, no apparent changes were detected in the hole-only device current density versus voltages (J-V) characteristics with and without drowning in ACN.

From the all-face-down SAM film, as shown schematically in Fig. 1c, we expect to obtain directional emission compared to that of the random-oriented one. Since CQWs possess nearly all in-plane transition dipole moments, their all-face-down ensemble film should act as an anisotropic emissive film. To realize the direction of emission ($\Theta$) of the deposited all-face-down single monolayer CQW film, we have used the back-focal plane (BFP) imaging technique. In this imaging, we found the ratio of the horizontal ($p_\parallel$) component to the sum of the horizontal and vertical ($p_\perp$) elements ($\Theta = p_\parallel / (p_\parallel + p_\perp)$), which experimentally quantified the directionality of TDMs. Fig. S4 illustrates the schematic of our BFP setup. In this experiment, spin-casted and single monolayer self-assembled films of NPLs on 200 μm thick quartz substrates were excited by a 400 nm laser, and a CCD camera took the back-focal plane images. Figs. 2a-b show the simulated intensity profile for fully in-plane (IP) and out-of-plane (OP) TDMs with the $k_x$-intensity diagrams for the p-polarized emission, given next to the images, respectively. The BFP images of self-assembled single monolayer and spin-coated samples with the $k_x$-intensity diagrams for the p-polarized emission are provided in figs. 2c-d. Using the intensity diagrams of fully IP, OP and their fit to the self-assembled film's intensity diagram (see fig. 2c), we found $\Theta$ equal to 92%. By repeating the same process for the random-oriented samples (see fig. 2d), we found the $\Theta$ equal to 67% (see the SI for the measurement and simulation details). Scott et al.[23] reported $\Theta$ was 95% for all-face-down CdSe NPLs and Shendre et al.[5] found $\Theta$ of 91% for all-face-down CdSe/CdS@CdZnS. Our measured $\Theta$ agrees with these results in the literature. These results confirm that in the all-face-down SAM film, a bright plane is directed toward the glass substrate.

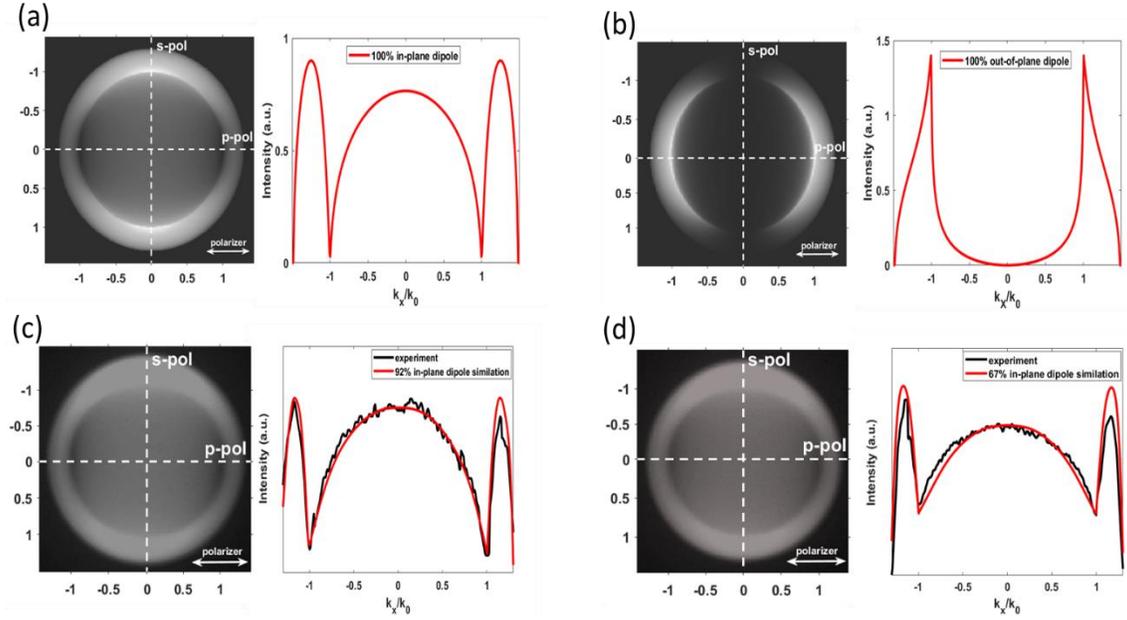

Fig. 2. Emission profile and intensity diagrams of a) fully IP TDMs, b) full OP TDMs, c) self-assembled monolayer film on a quartz substrate with 92% IP TDMs fitting curve, and d) spin-casted film on a quartz substrate with 67% IP TDMs curve.

Based on the measured Θ values, we calculated the light outcoupling efficiencies using the finite-discrete time-domain (FDTD) method by commercially available Lumerical FDTD solutions (ANSYS Inc) (for simulation details, see SI). Considering the HIL and HTLs structures, the calculated outcoupling efficiency for the device fabricated by an orientation-controlled film that has Θ = 92% was 34%. In contrast, it was 22% for the control sample made by spin-casting, which has Θ equal to 67%. Here, we computed a significant enhancement of 54.5% in the outcoupling efficiency.

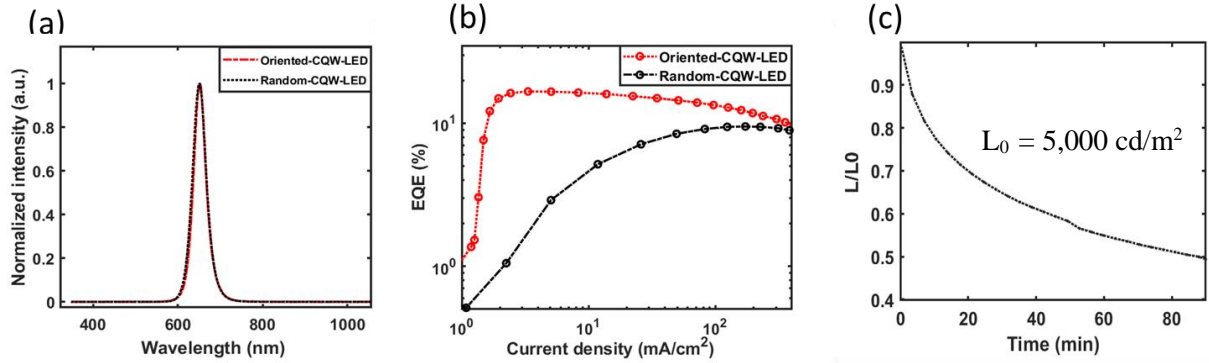

Fig. 3. a) Normalized electroluminescence (EL) intensity spectra for oriented-CQW-LED and random-CQW-LED devices. b) EQE versus current density for oriented-CQW-LED and random-CQW-LED devices. c) Stability data from the oriented-CQW-LED device at the starting luminance of 5,000 cd/m$^2$.

To study the performance of devices, we have fabricated CQW-LEDs by SAM film of HIS NPLs and spin-coated ones as a control sample. The normalized EL spectra (see fig. 3c) for the device fabricated using the all-face-down orienting self-assembly (oriented-CQW-LED) and randomly-orienting spin-casting (random-CQW-LED) film of NPLs show symmetric peaks at 650 nm without any shift in the peak position. Also, efficiency graphs (see fig. 3b) illustrate a peak EQE of 16.3% for oriented-CQW-LED and 9.5% for random-CQW-LED. By comparing two devices, we detected a 71.5% enhancement in EQE. Based on our calculations, 54.5% of this improvement resulted from the enhanced light-outcoupling, and the rest 17% is attributed to the facilitated charge injection and reduced reabsorption in the case of using an ordered single monolayer of CQWs. As shown in Fig. 3b, the oriented-CQW-LED device has reached its maximum EQE at lower current densities than the random-CQW-LED ones. This observation indicates the facilitated charge injection enabled by the self-assembly. Our devices are simply sealed with a glass coverslip and ultraviolet-curable resin, delivering good ambient stability. The half-lifetime ($T_{50}$), defined as the length of time for emitters to reach half of the initial brightness ($L_0/2$), for the oriented-CQW-LED device which was measured at the initial luminance of 5,000 cd/m$^2$, was $T_{50} = 89$ min (see fig. 3c). Using the formula of $L_0^n T_{50} = L_1^n T$[17], [36], [38], [50], [51] and assuming an acceleration factor of n = 1.5[17], we estimated the lifetime of 525 h at an initial luminance of 100 cd/m². Kelestemur et al.[15] reported a peak EQE of 9.92% for all-solution-processed CQW-LED using CdSe/CdS@CdZnS core/crown@HIS with a $T_{50}$ of 560 h, and Giovanella et al.[52] reported a device with a maximum EQE of 8.39% using CdSe/CdZnS core/HIS and a peak brightness of 1,500 cd/m². Thus far, our oriented-CQW-LED devices show the highest EQE among the all-solution processed CQWs. However, the maximum EQE value is still slightly lower than the record reported for CQW-LEDs[17].

To further push the efficiency of our devices, we changed the long native ligands to shorter ones. Since the OA and OLA ligands of our as-synthesized NPLs have a very long organic tail, their length limits the charge transport rate from the carrier transport layer (CTL) to the EML, enlarging the electron-hopping distance and creating a potential barrier for energy transfer. Thus, we changed the native long OA and OLA ligands with the shorter 2-ethylhexane-1-thiol (EHT) to reduce this energy barrier for the charge transport. Song et al.[38] reported a high EQE of 30.9% for red-QLED using the EHT ligands on colloidal quantum dots.

To confirm the ligand-exchange[49], [53], we used the Fourier-transform infrared spectroscopy (FTIR) (fig. S3). In the FTIR spectrum, the bending vibrations of the amine group of OLA (located at 1,641cm$^{-1}$) of the HIS NPLs have vanished after the ligand exchange (LE-HIS samples). We found out that the ligand-exchange process affects the electrical properties of NPLs without changing the optical properties. The photoluminescence (PL) and absorption spectra for the HIS and LE-HIS samples remained unchanged (see fig. 4a). However, the main effect of the switch of the ligands was on the valance band position of NPLs. We measured the band offsets using the XPS technique. Here, the XPS analysis of energy bands demonstrates the effects of ligands dipole. Fig. 4b illustrates the proximity of the measured $E_F$-$E_{VBM}$ for NPLs with OA (0.93 eV) and EHT

(1.26 eV), suggesting that the ligand exchange barely perturbs the Fermi level of NPLs. However, the main difference has been detected in the secondary electron cut-off region, indicating the ligands' dipole effect. Using two cut-off regions (fig. 4b) and the method of Miller et al.[54] and considering corrections[55], the calculated $E_{VBM}$ values were ~5.93 for the LE-HIS samples and ~6.61 eV for the HIS samples. Since the highest occupied molecular orbital (HOMO) of our used PVK was 5.8 eV, the potential barrier for the holes' movement from HTL to the NPLs is reduced from 0.87 eV to 0.13 eV after the ligand-exchange process.

Moreover, the EHT ligands preserve the high-PLQY of the emitters. In our case, the PLQY of ensemble NPLs remained as high as 88% after the ligand exchange. Even after six times washing, the PLQY decreased only to 85%. This observation confirms the robust attachment of ligands to the NPLs by the thiol group[38]. Also, the PLQY of NPLs with and without the ligand exchange in several substrates remained close to each other (see fig. 4a). Furthermore, since the EHT ligands have a shorter length than the native ligands, the SAM film of LE-HIS NPLs is denser than the HIS NPLs. We have calculated $29 \times 10^8$ NPLs/mm$^2$ and $24 \times 10^8$ NPLs/mm$^2$ for the cases of LE-HIS and HIS, respectively.

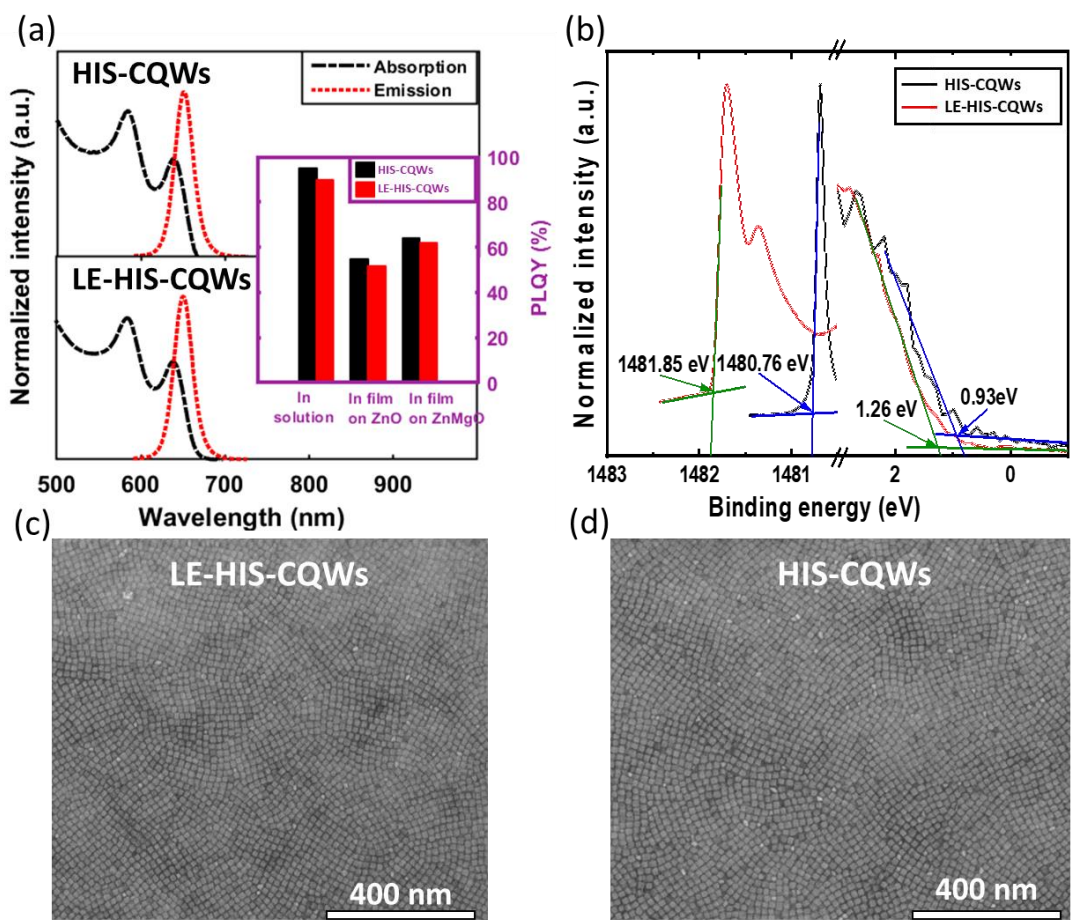

Fig. 4. a) Absorption and photoluminescence (PL) spectra of the hot-injection shell (HIS) grown (top) and ligand-exchanged (LE-HIS) (bottom) CQWs. Inset, PLQY of HIS-CQWs and LE-HIS-CQWs in solution, on ZnO film and $Zn_{0.95}Mg_{0.05}O$ film. b) XPS data of second cut-off region (left-hand side) and valance band region (right-hand side). Sem images of self-assembled monolayer of c) LE-HIS-CQWs and d) HIS-CQWs on a substrate coated with the layers of ITO/PEDOT:PSS/p-TPD/PVK.

The normalized EL spectrum (see fig. 5a) of devices fabricated use of the SAM film of LE-HIS NPLs (oriented-LE-CQW-LED) exhibits a symmetric peak at 651 nm, corresponding to a CIE coordinate of (0.710,0.289) (see fig. 5b), which is relatively close to the spectral locus of (0.708, 0.292) known as saturated red color. This CIE is ideal for new generation ultra-high-definition television (UHDTV) applications[17], [38], [50]. The inset photograph is taken from the random-LE-CQW-LED device operating at 5.0 V. Current density-voltage and luminance-voltage (J-V-L) diagrams (see fig. 5c) exhibit an abrupt increase in current and luminance once they reach the sub-bandgap turn-on voltage of ~1.7 V at a current density of 1 mA/cm², showing a typical diodic behavior with a maximum brightness of 19,800 cd/m² at 8.0 V. The peak EQE value of 18.1% at a current density of ~2.36 mA/cm² has been measured for these devices (see fig. 5d). This value is the highest reported EQE for all-solution processed CQW-LEDs thus far. The $J_{90}$, the current density in which EQE drops by 10%, is 95.58 mA/cm², which indicates outstanding stability and low-efficiency roll-off. Also, the histogram of 35 devices exhibits an average peak EQE value of 16.8%, suggesting suitable device reproducibility. Using the LE-HIS film of NPLs, we observed an 11% improvement in EQE compared to the HIS NPLs. Moreover, these devices of LE-HIS CQWs show outstanding color stability of different luminance values (see fig. 5e). Fig. 5f shows the emission pattern of the devices of random-LE-CQW-LED, oriented-LE-CQW-LED, and a perfect Lambertian emission. The oriented-LE-CQW-LED's angle of emission is narrower than the random-LE-CQW-LED, and both are lower than the ideal Lambertian emission: the directionality in emission is preserved in the oriented-LE-CQW-LED.

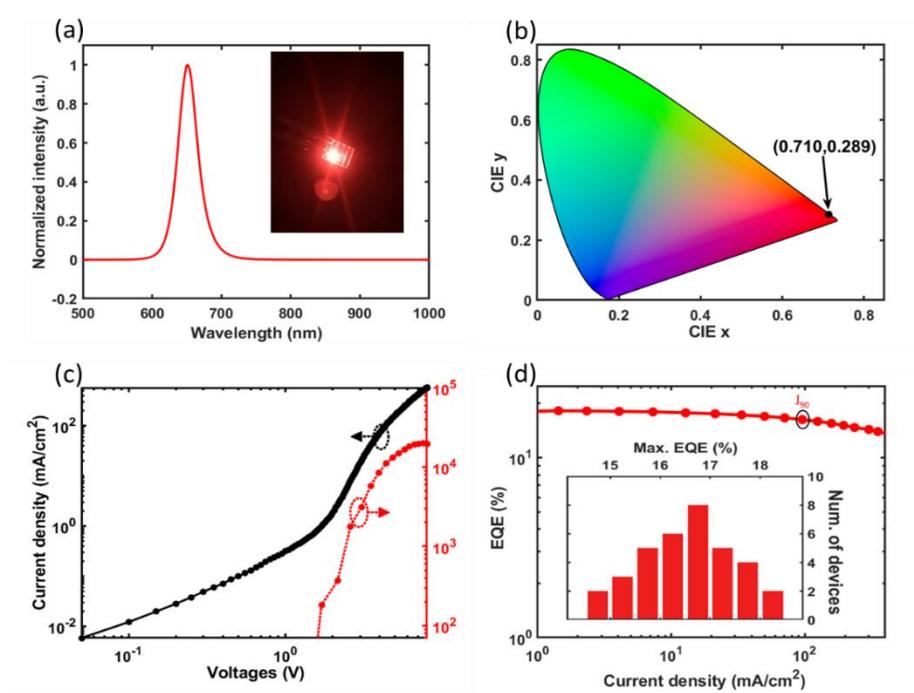

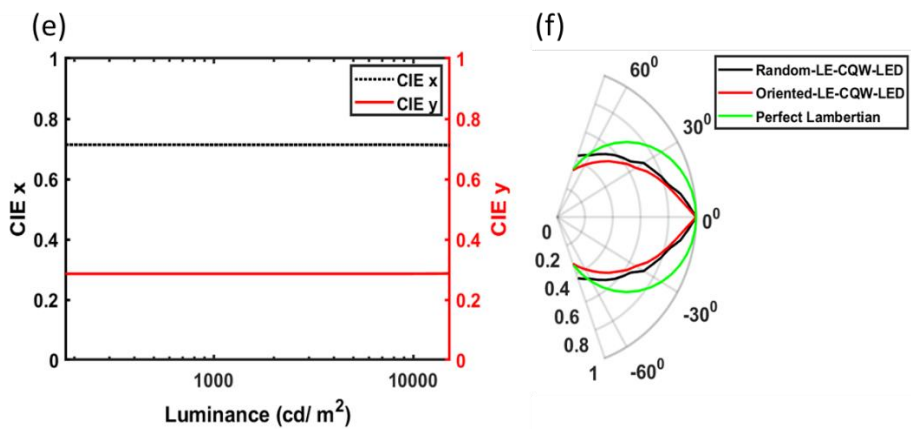

Fig. 5. a) EL spectra of the LED fabricated by self-assembly and ligand exchange (oriented-LE-CQW-LED). Inset: a photograph of the device operating at 5.0 V. b) Corresponding CIE coordinates. c) Current density vs. voltage and luminance vs. voltages diagrams of oriented-LE-CQW-LED. d). EQE vs. current density of (oriented-LE-CQW-LED) with $J_{90}$. Inset maximum EQE histogram of oriented-LE-CQW-LED measured from 35 devices. e) CIE coordinate as a function of the luminance. f) Emission pattern of the oriented-LE-CQW-LED, random-LE-CQW-LED devices, and that for simulated perfect Lambertian emission.

Our devices were simply sealed with a glass coverslip and ultraviolet-curable resin, delivering good ambient stability. The half-lifetime, measured at the initial luminance of 5,000 cd/m², illustrates a $T_{50}$ = 42 min (see fig. 6a). Using the formula of $L_0{}^n T_{50} = L_1{}^n T$[17], [36], [38], [50], [51] and assuming an acceleration factor of n = 1.5, we estimate the lifetime of 247 h at an initial luminance of 100 cd/m². Our reported lifetime is nineteen-fold higher than the record organic-inorganic CQW-LEDs[17], but it is lower than the oriented-CQW-LED devices. The accumulation of charges in the EML due to shorter ligands and heating issues are considered the main reasons for this lower $T_{50}$.

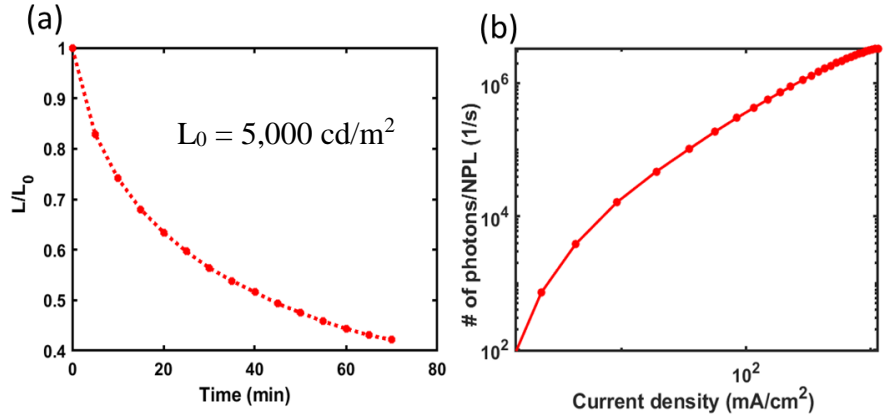

Fig. 6. a) Stability data from the oriented-LE-CQW-LED device at the starting luminance of 5,000 cd/m$^2$. b) Number of outcoupled photons per NPL vs. current density in oriented-LE-CQW-LED.

Moreover, the self-assembled monolayer film of NPLs enables us to explore the number of generated photons per individual NPL. Since the liquid-air self-assembly yields a highly uniform ensemble film, we were able to calculate the number of NPLs inside the active region of the devices. Using SEM images (see fig. 1c) and image analysis software (ImageJ), we have found ~13×10$^9$ NPLs inside the device's active area (4.5 mm$^2$). Fig 6b shows the diagram of the number of outcoupled photons per NPL versus current density. At the maximum EQE, we have ~2,180 photon/NPL per unit time. Considering the exciton lifetime of ~10 ns in NPLs [17], [40], we have calculated ~2×10$^{-5}$ photon/NPL per unit time. However, this is the number of outcoupled photons, and to calculate the exciton number, we have computed the internal quantum efficiency (IQE) using the equation of EQE = IQE×$\eta_{out}$[17]. Since the $\eta_{out}$ is 34%, we have found ~6×10$^{-5}$ exciton/NPL at a given time. This result shows that we are in the single excitonic regime in our LEDs. Also, this value at the maximum current density is ~9×10$^{-2}$ exciton/NPL, which still is in the single exciton regime. It has been reported that for an amplified spontaneous emission (ASE), 2-5 exciton/NPL is required[10], which is two orders of magnitude higher than the current density achieved in this study.

In conclusion, we have shown highly efficient all-solution-processed CQWs with record EQE in the saturated red with excellent stability and a long lifetime. Such an impressive electroluminescence performance results from exploiting the orientation-controlled single monolayer film of NPLs in the EML of devices. This film enables a 71.5% enhancement in the EQE compared to the random-oriented (spin-casted) one. Also, this film yields the lowest possible surface roughness among all other deposition methods and enhances the charge injection efficiencies. In addition, changing the native OA and OLA ligands with EHT leads to a densely packed film of NPLs and improves the charge injection rate (11%) by shortening the hopping distance. By implementing the orientation-controlled film of ligand-exchanged NPLs in the devices, we demonstrated a high EQE value of 18.1%, with a peak brightness of 19,800 cd/m$^2$ in an extremely saturated red color at a CIE coordinate of (0.710, 0.289) with a $T_{50}$ of 247 h and an impressive $J_{90}$ of 95.58 mA/cm$^2$. Our findings demonstrate the applicability of our single monolayer oriented-CQW LED architecture to fabricate high-performance all-solution processed normal, flexible, and large-area CQW-LEDs.

**Supporting Information**

Supporting Information is available from the Wiley Online Library or from the author.

**Acknowledgements**

The authors gratefully acknowledge the financial support in part from Singapore National Research Foundation under the programs of NRF-NRFI2016-08, NRF-CRP14-2014-03 and the Science and Engineering Research Council, Agency for Science, Technology and Research (A*STAR) of Singapore and in part from TUBITAK 115F297, 117E713, 119N343, and 20AG001. B.C. acknowledges TUBITAK 2218. H. V. D. also acknowledges the support from TUBA.

**References**

[1] C. B. Murray, D. J. Norris, and M. G. Bawendi, "Synthesis and Characterization of Nearly Monodisperse CdE (E = S, Se, Te) Semiconductor Nanocrystallites," *J. Am. Chem. Soc.* **1993**, vol. 115, no. 19, pp. 8706–8715.

[2] S. Volk, N. Yazdani, and V. Wood, "Manipulating Electronic Structure from the Bottom-Up: Colloidal Nanocrystal-Based Semiconductors," *J. Phys. Chem. Lett.* **2020**, vol. 11, no. 21, pp. 9255–9264.

[3] S. Ithurria and D. V. Talapin, "Colloidal Atomic Layer Deposition (c-ALD) using self-limiting reactions at nanocrystal surface coupled to phase transfer between polar and nonpolar media," *J. Am. Chem. Soc.* **2012**, vol. 134, no. 45, pp. 18585–18590.

[4] S. Ithurria, M. D. Tessier, B. Mahler, R. P. S. M. Lobo, B. Dubertret, and A. L. Efros, "Colloidal nanoplatelets with two-dimensional electronic structure" *Nat. Mater.* **2011**, vol. 10, no. 12, pp. 936–941.

[5] S. Shendre, S. Delikanli, M. Li, D. Dede, Z. Pan, S. T. Ha, Y. H. Fu, P. L. Hernandez-Martinez, J. Yu, O. Erdem, A. I. Kuznetsov, C. Dang, T. C. Sum, and H. V. Demir, "Ultrahigh-efficiency aqueous flat nanocrystals of CdSe/CdS@Cd1−xZnxS colloidal core/crown@alloyed-shell quantum wells," *Nanoscale* **2019**, vol. 11, no. 1, pp. 301–310.

[6] Z. Chen, B. Nadal, B. Mahler, H. Aubin, and B. Dubertret "Quasi-2D Colloidal Semiconductor Nanoplatelets for Narrow Electroluminescence," *Adv. Funct. Mater.* **2014**, vol. 24, pp. 295–302.

[7] E. Lhuillier, S. Pedetti, S. Ithurria, B. Nadal, H. Heuclin, and B. Dubertret, "Two-

Dimensional Colloidal Metal Chalcogenides Semiconductors: Synthesis, Spectroscopy, and Applications," *Acc. Chem. Res.* **2015**, 48, no. 1, pp. 22–30.

[8]     A. G. Vitukhnovsky, V. S. Lebedev, A. S. Selyukov, A. A. Vashchenko, R. B. Vasiliev, and M. S. Sokolikova, "Electroluminescence from colloidal semiconductor CdSe nanoplatelets in hybrid organic-inorganic light emitting diode," *Chem. Phys. Lett.* **2015**, vol. 619, pp. 185–188.

[9]     S. Delikanli, O. Erdem, F. Isik, H. D. Baruj, F. Shabani, H. B. Yagci, E. G. Durmusoglu, and H. V. Demir, "Ultrahigh Green and Red Optical Gain Cross Sections from Solutions of Colloidal Quantum Well Heterostructures," *J. Phys. Chem. Lett.* **2021**, vol. 12, no. 9, pp. 2177–2182.

[10]    S. Delikanli, F. Isik, F. Shabani, H. D. Baruj, N. Taghipour, and H. V. Demir, "Ultralow Threshold Optical Gain Enabled by Quantum Rings of Inverted Type-I CdS / CdSe Core / Crown Nanoplatelets in the Blue," *Adv. Opt. Mater*. **2021**, no. 9, vol. 2002220, pp. 1-6.

[11]    J. Maskoun, N. Gheshlaghi, F. Isik, S. Delikanli, Em Y. Erdem, and H. V. Demir, "Optical Microfluidic Waveguides and Solution Lasers of Colloidal Semiconductor Quantum Wells," *Adv. Mater.* **2021**, vol. 2007131, pp. 1–6.

[12]    N. Gheshlaghi, S. Foroutan-Barenji, O. Erdem, Y. Altintas, F. Shabani, M. H. Humayun, and H. V. Demir, "Self-Resonant Microlasers of Colloidal Quantum Wells Constructed by Direct Deep Patterning," *Nano Lett.* **2021**, 21, 11, 4598–4605.

[13]    Y. Shirasaki, G. J. Supran, M. G. Bawendi, and V. Bulović, "Emergence of colloidal quantum-dot light-emitting technologies," *Nat. Photon.* **2013**, vol. 7, no. 1, pp. 13–23,.

[14] M. Pelton, "Carrier Dynamics, Optical Gain, and Lasing with Colloidal Quantum Wells," *J. Phys. Chem*. C **2018**, 122, 20, 10659–10674.

[15] Y. Kelestemur, Y. Shynkarenko, M. Anni, S. Yakunin, M. L. De Giorgi, and M. V Kovalenko, "Colloidal CdSe Quantum Wells with Graded Shell Composition for Low-Threshold Amplified Spontaneous Emission and Highly Efficient Electroluminescence," *ACS Nano* **2019**, 13, 12, 13899–13909.

[16] A. Polovitsyn, Z. Dang, J. L. Movilla, B. Martin-Garcia, A. H. Khan, G. H. V. Bertrand, R. Brescia, and I. Moreels, "Synthesis of Air-Stable CdSe/ZnS Core−Shell Nanoplatelets with Tunable Emission Wavelength," *Chem. Mater*. **2017**, 29, 13, 5671–5680.

[17] B. Liu, Y. Altintas, L. Wang, S. Shendre, M. Sharma, H. Sun, E. Mutlugun, and H. V. Demri, "Record High External Quantum Efficiency of 19.2% Achieved in Light-Emitting Diodes of Colloidal Quantum Wells Enabled by Hot-Injection Shell Growth," *Adv.Mater*. **2020**, no. 32, pp. 1905824.

[18] A. A. Rossinelli, H. Rojo, A. S. Mule, M. Aellen, A. Cocina, E. D. Leo, R. Schaublin, and D. J. Norris, "Compositional Grading for Efficient and Narrowband Emission in CdSe-Based Core/Shell Nanoplatelets," *Chem. Mater*. **2019**, no. 31, vol. 22, pp. 9567–9578.

[19] M. Sharma, S. Delikanli, and H. V. Demir, "Two-Dimensional CdSe-Based Nanoplatelets: Their Heterostructures, Doping, Photophysical Properties, and Applications," *Proc. IEEE*, **2020,** vol. 108, no. 5, pp. 655–675.

[20] X. Dai, Z. Zhang, Y. Jin, Y. Niu, H. Cao, X. Liang, L. Chen, J. Wang, and X. Leng, "Solution-processed, high-performance light-emitting diodes based on quantum dots," *Nature*, vol. 515, no. 7525, pp. 96–99, **2014**.

[21] S. Noway, B. C. Krummacher, J. Frischeisen, N. A. Reinke, and W. Brutting, "Light extraction and optical loss mechanisms in organic light-emitting diodes : Influence of the emitter quantum efficiency," *J. Appl. Phys.* **2008**, no. 104, pp. 123109.

[22] J. A. Schuller, S. Karaveli, T. Schiros, K. He, S. Yang, I. Kymissis, J. Shan, and R. Zia, "Orientation of luminescent excitons in layered nanomaterials," *Nat. Nanotechnol.* **2013**, vol. 8, no. 4, pp. 271–276.

[23] R. Scott, J. Heckmann, A. V. Prudnikau, A. Antanovich, A. Mikhailov, N. Owschimikow, M. Artemyev, J. I. Climente, U. Woggon, N. B. Grosse, and A. W. Achtstein, "Directed emission of CdSe nanoplatelets originating from strongly anisotropic 2D electronic structure," *Nat. Nanotechnol.* **2017**, vol. 12, no. 12, pp. 1155–1160.

[24] M. Sharma, K. Gungor, A. Yeltik, M. Olutas, B. Guzelturk, Y. Kelestemur, T. Erdem, S. Delikanli, J. R. McBride, and H. V. Demir, "Near-Unity Emitting Copper-Doped Colloidal Semiconductor Quantum Wells for Luminescent Solar Concentrators," *Adv. Mater.* **2017**, vol. 29, no. 30, pp. 1–10.

[25] J. Heckmann, R. Scott, A. V. Prudnikau, A. Antanovich, N. Owschimikow, M. Artemyev, J. I. Climente, U. Woggon, N. B. Grosse, and A. W. Aschtstein, "Directed Two-Photon Absorption in CdSe Nanoplatelets Revealed by k-Space Spectroscopy," *Nano Lett.* **2017**, vol. 17, no. 10, pp. 6321–6329.

[26] Y. Gao, M. C. Weidman, and W. A. Tisdale, "CdSe Nanoplatelet Films with Controlled Orientation of their Transition Dipole Moment," *Nano Lett.* **2017**, vol. 17, no. 6, pp. 3837–3843.

[27] A. W. Achtstein, A. Antanovich, A. Prudnikau, R. Scott, U. Woggon, and M. Artemyev,

"Linear Absorption in CdSe Nanoplates: Thickness and Lateral Size Dependency of the Intrinsic Absorption," *J. Phys. Chem.* **2015**, *C*, vol. 119, no. 34, pp. 20156–20161.

[28] A. B. Vasista, D. K. Sharma, and G. V. P. Kumar, "Fourier Plane Optical Microscopy and Spectroscopy," *Digit. Encycl. Appl. Phys.* **2019**, pp. 1–14.

[29] X. Chen, C. Ji, Y. Xiang, X. Kang, B. Shen, and T. Yu, "Angular distribution of polarized light and its effect on light extraction efficiency in AlGaN deep-ultraviolet light-emitting diodes," *Opt. Express*, **2016**, vol. 24, no. 10, p. A935,.

[30] M. A. Lieb, J. M. Zavislan, and L. Novotny, "Single-molecule orientations determined by direct emission pattern imaging," *J. Opt. Soc. Am. B*, **2004**, vol. 21, no. 6, p. 1210.

[31] H. Budde, N. Coca-Lopez, X. Shi, R. Ciesielski, A. Lombardo, D. Yoon, A. C. Ferrari, A. Hartschuh, "Raman Radiation Patterns of Graphene," *ACS Nano*, **2016**, vol. 10, no. 2, pp. 1756–1763.

[32] P. Bai, A. Hu, Y. Liu, Y. Jin, and Y. Gao, "Printing and in Situ Assembly of CdSe/CdS Nanoplatelets as Uniform Films with Unity In-Plane Transition Dipole Moment," *J. Phys. Chem. Lett.* **2020**, vol. 11, no. 11.

[33] A. Lefrançois, B. Luszczynska, B. Pepin-Donat, C. Lombard, B. Bouthinon, J. Verilhac, M. Gromova, J. Faure-Vincent, S. Pouget, F. Chandezon, S. Sadki, and P. Reiss, "Enhanced charge separation in ternary P3HT/PCBM/CuInS2 nanocrystals hybrid solar cells," *Sci. Rep.* **2015**, vol. 5, no.7768.

[34] O. Erdem, K. Gungor, B. Guzelturk, I. Tanriover, M. Sak, M. Olutas, D. Dede, Y. Kelestemur, and H. V. Demir, "Orientation-Controlled Nonradiative Energy Transfer to

Colloidal Nanoplatelets: Engineering Dipole Orientation Factor," *Nano Lett.* **2019**, vol. 19, no. 7, pp. 4297–4305.

[35] Q. Yuan, T. Wang, P. Yu, H. Zhang, H. Zhang, and W. Ji, "A review on the electroluminescence properties of quantum-dot light-emitting diodes," *Org. Electron.* **2021**, vol. 90, no. p. 106086.

[36] B. Liu, S. Delikanli, Y. Gao, D. Dede, K. Gungor, and H. V. Demir, "Nanocrystal light-emitting diodes based on type II nanoplatelets," *Nano Energy* **2018**, vol. 47, no.106086 pp. 115–122.

[37] H. Shen, W. Cao, N. T. Shewmon, C. Yang, L. S. Li, and J. Xue, "High-Efficiency, Low Turn-on Voltage Blue-Violet Quantum-Dot-Based Light-Emitting Diodes," *Nano Lett* **2015**, vol. 20, p. 47.

[38] J. Song, O. Wang, H. Shen, Q. Lin, Z. Li, and L. Wang, "Over 30 % External Quantum Efficiency Light-Emitting Diodes by Engineering Quantum Dot-Assisted Energy Level Match for Hole Transport Layer." *Adv. Funct. Mater*. **2019**, 29, 1808377.

[39] C. Pu, X. Dai, Y. Shu, M. Zhu, Y. Deng, Y. Jin, and X. Peng, "Electrochemically-stable ligands bridge the photoluminescence-electroluminescence gap of quantum dots," *Nat. Commun.* **2020**, vol. 11, no. 1, pp. 1–10.

[40] Y. Altintas, K. Gungor, Y. Gao, M. Sak, U. Quliyeva, G. Bappi, E. Mutlugun, E. H. Sargent, and H. V. Demir, "Giant Alloyed Hot Injection Shells Enable Ultralow Optical Gain Threshold in Colloidal Quantum Wells," *ACS Nano* **2019**, vol. 13, no. 9, pp. 10662–10670.

[41]  J. M. Pietryga, Y. Park, J. Lim, A. F. Fidler, W. K. Bae, S. Brovelli, and V. I. Klimov, "Spectroscopic and device aspects of nanocrystal quantum dots," *Chem. Rev.* **2016**, vol. 116, no. 18, pp. 10513–10622.

[42]  Y. Altintas, U. Quliyeva, K. Gungor, O. Erdem, Y. Kelestemur, E. Mutlugun, M. V. Kovalenko, H. V. Demir, "Highly Stable, Near-Unity Efficiency Atomically Flat Semiconductor Nanocrystals of CdSe/ZnS Hetero-Nanoplatelets Enabled by ZnS-Shell Hot-Injection Growth," Small **2019**, 15, 1804854.

[43]  J. Yu, M. Sharma, Y. Wang, S. Delikanli, H. D. Baruj, A. Sharma, H. V. Demir, and C. Dang, "Modulating Emission Properties in a Host–Guest Colloidal Quantum Well Superlattice," *Adv. Opt. Mater.* **2021**, vol. 2101756, p. 2101756.

[44]  Y. Sun, Y. Jiang, H. Peng, J. Wei, S. Zhang, and S. Chen, "Efficient quantum dot light-emitting diodes with a Zn 0.85 Mg 0.15 O interfacial modification layer," *Nanoscale*, **2017**, vol. 9, pp. 8893–9248,.

[45]  H. Zhang, Q. Su, and S. Chen, "Quantum-dot and organic hybrid tandem light-emitting diodes with multi-functionality of full-color-tunability and white-light-emission." *Nat. Commun.* **2020**, 11, 2826

[46]  Q. Su, H. Zhang, and S. Chen, "Flexible and tandem quantum-dot light-emitting diodes with individually addressable red/green/blue emission." *npj Flexible Electronics* **2021**, 5, 8.

[47]  F. Shabani, H. D. Baruj, I. Yurdakul, S. Delikanli, N. Gheshlaghi, F. Isik, B. Liu, Y. Altintas, B. Canımkurbey, and H. V. Demir, "Deep-Red-Emitting Colloidal Quantum Well Light-Emitting Diodes Enabled through a Complex Design of Core/Crown/Double

Shell Heterostructure," *Small* **2021**, vol. 2106115, p. 2106115.

[48] L. Qian, Y. Zheng, J. Xue, and P. H. Holloway, "Stable and efficient quantum-dot light-emitting diodes based on solution-processed multilayer structures," *Nat. Photon.* **2011**, 5, 543-548.

[49] Z. Li, Y. Hu, H. Shen, Q. Lin, L. Wang, H. Wang, W. Zhao, and L. S. Li., "Efficient and long-life green light-emitting diodes comprising tridentate thiol capped quantum dots," *Laser Photonics Rev*. **2017**, vol. 11, no. 1, p. 1600227.

[50] X. Dai, Z. Zhang, Y. Jin, Y. Niu, H. Cao, X. Liang, L. Chen, J. Wang, and X. Peng "Solution-processed, high-performance light-emitting diodes based on quantum dots," *Nature* **2014**, 515, 96-99.

[51] H. Volkan, S. Nizamoglu, T. Erdem, E. Mutlugun, N. Gaponik, and A. Eychmüller, "Quantum dot integrated LEDs using photonic and excitonic color conversion," *Nano Today* **2011**, vol. 6, no. 6, pp. 632–647.

[52] U. Giovanella, M. Pasini, M. Lorenzon, F. Galeotti, C. Lucchi, F. Meinardi, S. Luzzati, B. Dubertret, and S. Brovelli, "Efficient Solution-Processed Nanoplatelet-Based Light-Emitting Diodes with High Operational Stability in Air," *Nano Lett* **2018**, vol. 18, p. 24.

[53] H. Shen, W. Cao, N. T. Shewmon, C. Yang, L. S. Li, and J. Xue, "High-Efficiency, Low Turn-on Voltage Blue-Violet Quantum-Dot-Based Light-Emitting Diodes," *Nano Lett* **2015**, vol. 6, p. 57.

[54] E. M. Miller, D. M. Kroupa, J. Zhang, P. Schulz, A. R. Marshall, A. Kahn, S. Lany, J. M. Luther, M. C. Beard, C. L. Perkins, and J. V. D. Lagemaat, "Revisiting the Valence and

Conduction Band Size Dependence of PbS Quantum Dot Thin Films," *ACS Nano* **2016**, vol. 10, p. 2021.

[55] D. M. Kroupa, M. Vörös, N. P. Brawand, B. W. McNichols, E. M. Miller, J. Gu, A. J. Nozik, A. Sellinger, G. Galli, and M. C. Beard "Tuning colloidal quantum dot band edge positions through solution-phase surface chemistry modification," *Nat. Commun.* **2017**, vol. 8, no. May, pp. 2–9.

**Supporting information**

**Highly-directional, highly-efficient solution-processed light-emitting diodes of all-face-down oriented colloidal quantum wells**

*Hamed Dehghanpour Baruj, Iklim Yurdakul, Betul Canimkurbey, Ahmet Tarik Isik, Farzan Shabani, Savas Delikanli, Sushant Shendre, Onur Erdem, Furkan Isik, and Hilmi Volkan Demir[*]*

**Materials:** Cadmium acetate dihydrate (Cd(Ac)$_2$×2H$_2$O, 98%), cadmium nitrate tetrahydrate (≥99.0%), cadmium acetate (Cd(Ac)$_2$, 99.995%), zinc-acetate dihydrate (Zn(Ac)$_2$.2H$_2$O, 98%), zinc-acetate (Zn(Ac)$_2$, 99.99%), sodium myristate (≥99.0%), sulfur (99.98%), ammonium sulfide solution ((NH4)2S, 40-48 wt. % in H$_2$O), selenium (99.99%), oleic acid (OA, 90%), oleylamine (OLA, 70%), 1- Octanethiol (≥98.5%), 1-octadecene (ODE, 90%), n-hexane (≥97.0%), ethanol (absolute), methanol (≥99.7%), toluene (≥99.5%), acetonitrile (ACN, 99.9 %), poly(9-vinylcarbazole) (PVK), magnesium acetate tetrahydrate (≥99%), tetramethylammonium hydroxide (TMAH, 98%), chlorobenzene (>98%), m-xylene (>98%), 1-4 dioxane (>99%), 2-ethylhexane-1-thiol (EHT) were obtained from Sigma Aldrich. Poly (N, N′-bis(4-butylphenyl)-N,N′-bis(phenyl)-benzidine) (p-TPD) was purchested from Lumtec. Poly (ethylene dioxythiophene): polystyrene sulfonate (PEDOT: PSS) was obtained from Osilla (AI 4083). All chemicals were used without further modification.

**Methods:**

**Synthesis of cadmium myristate**
Cadmium myristate was synthesized according to the previous recipes in the literature, with slight modification.[1] In a typical synthesis, 2.46 g of cadmium nitrate tetrahydrate and 6.26 g of sodium myristate were dissolved in 80 and 500 mL of methanol, respectively. After 3 h stirring, they were mixed and kept stirring for 5 h. Then, a white color cadmium myristate was precipitated by centrifugation. To purify and remove the impurities, the cadmium myristate powder was washed two times using methanol. Finally, the product was kept overnight under vacuum at room temperature and stored under ambient conditions.

**Synthesis of 4 ML CdSe core nanoplatelets (NPLs)**
4 monolayer (ML) CdSe NPLs were synthesized according to a reported protocol. First, 24 mg of Se powder and 340 mg of cadmium myristate, and 30 mL of ODE were loaded into a 100 mL three-neck flask and degassed at 95 °C for 1 h. Then, under argon gas, the temperature was set to 240 °C. When the color of the solution changed to an orangish color around 195 °C, 120 mg of cadmium acetate dihydrate was added to the reaction. The solution was kept stirring at 240 °C for 8 min. Finally, the growth was terminated by adding 1.5 mL OA and quenched in a cold-water bath. The 4 ML CdSe NPLs purified through size-selective precipitation to obtain a monodisperse solution. The final product was redispersed in hexane and kept for further use.

**Synthesis of CdSe/CdZnS core/ hot-injection shell (HIS) NPLs**

The core/HIS NPLs were synthesized based on published protocols with slight modification. The 4 ML solution in hexane, Zn-acetate (55 mg), Cd-acetate (22.5 mg), ODE (7.5 mg) were added to a 25 mL flask and degassed for 1 h at ambient temperature and 30 min at 80 °C. Then, under argon gas, 1 mL of OLA was added to the solution, and the temperature was set to 300 °C. Starting from 165 °C, the mixture of ODE (5 mL) and octanethiol (140 µL) was injected at a rate of 10 mL/h. Note that, after reaching 240 °C, the injection rate was decreased to 4 mL/h. After finishing the injection, the solution was kept for 1 h to complete the growth. The reaction was terminated by quenching with a cold-water bath. Then by adding 5 mL hexane and 3 mL of ethanol, the solution has been washed. Finally, the NPLs redispersed in toluene and kept for the next step.

**Ligand exchange of NPLs**
Under argon flows, 150 µL of 2-ethylhexane-1-thiol (EHT) ligands were added to 1 ML of NPLs in toluene solution (100 mg/mL). The solution was kept stirring for 2 h under the gas. Then, NPLs were precipitated by adding ethanol and redispersed in toluene. The process was repeated two more times to have a complete exchange, but in the second and third, 50 µL of EHT was added. Finally, the resultantly NPLs were washed by adding ethanol and redispersed in hexane.

**Synthesis of $Zn_{0.95}Mg_{0.05}O$**
To synthesize $Zn_{0.95}Mg_{0.05}O$, 2.95 mmol Zn-acetate dihydrates and 0.05 mmol magnesium acetate tetrahydrate were dissolved in 30 mL of dimethyl sulfoxide solution stirred vigorously (at 1,000 rpm). Then, the mixture of 5.5 mmol tetramethylammonium hydroxide in 10 mL ethanol was injected into the Zn-acetate solution at a rate of 50 mL/h. Under the ambient conditions, the mixture was kept being stirred for 2 h. Afterward, the $Zn_{0.95}Mg_{0.05}O$ nanocrystals (NCs) were precipitated by adding ethyl acetate and redispersed in ethanol. To improve $Zn_{0.95}Mg_{0.05}O$ nanoparticles solubility, inside the nitrogen-filled glovebox, 200 µL of ethanolamine was added to the mixture and kept stirring for 2 h. Finally, the obtained $Zn_{0.95}Mg_{0.05}O$ nanoparticles were washed by adding ethyl acetate and redispersed in ethanol.

**Device fabrication**
First, pre-patterned indium tin oxide (ITO) coated substrates were cleaned sequentially using detergent, distilled water, acetone, and isopropanol, for 15 min each. Then, the substrates were treated by ozone-plasma for 15 min in 30 W to completely clean the remained organics and uplift the work-function of ITO. Subsequently, PEDOT:PSS/isopropanol (1:1) solution (Osilla Al 4083, filtered through 0.25 µm PTFE membrane) were spin-casted on the patterned ITO-coated glass substrates at 3,000 rpm. for 60 s and baked at 120°C for 30 min. The coated substrates were transferred to a nitrogen-filled glove box ($O_2$<0.1 & $H_2O$<0.1 ppm). The substrates baked for 5 min in 110°C inside the glove box to remove the possible absorbed $H_2O$ while transferring. Poly-TPD (8 mg/mL in chlorobenzene), PVK (2 mg/mL in 1-4 dioxane), NPLs (15 mg/mL in hexane), $Zn_{0.95}Mg_{0.05}O$ (25 mg/mL in ethanol) were deposited sequentially by spin-casting at 2000 rpm for

60 s. The poly-TPD and PVK were baked at 110 and 150 °C for 30 min, respectively. NPLs and ZnMgO were baked at 60 °C for 20 min and 90 °C for 30 min, respectively. Finally, 100 nm of Al was deposited through a shadow mask by an oxygen-free thermal evaporation system (~$9\times10^{-7}$ torr) to create a 4.5 mm$^2$ active area. Finally, the devices were encapsulated by glass coverslip and UV-curable resin.

**Characterization**

We used Agilent Technologies (U3606A) electrometer to measure current-voltage characterization and an integrating sphere (Newport 5.3") coupled with Ocean Optics (QEPro) spectrometer for output light measurements. In the measurements, the devices were put in close contact with the input aperture of the sphere (they were not inside the sphere). Also, the hole size was relatively larger than the device's active area (from the light-emitting diode (LED) to the integration sphere), suggesting that the coupling factor for the emitted photons in the forward direction was unity. We have used a Konica Minolta cs-2000 spectroradiometer for the luminance intensity and $T_{50}$ measurements. Note that the $T_{50}$ data have been taken manually using spectroradiometers software (CS-S10W). The angular emission intensity (goniometric) measurements were carried out using a calibrated Si photodiode (Newport).

We used a fluorescence spectrophotometer – Cary Eclipse for photoluminescence and absorption spectra, PSIA Training instrument for atomic force microscopy (AFM) analysis, Thermo Fisher K-alpha XPS instrument for X-ray photoelectron spectroscopy (XPS) measurements, FEI focused ion beam (FIB)/scanning electron microscopy (SEM) system for microstructure analysis. Ellipsometric measurements were carried out by J. A. Woollam ellipsometer (V-VASE). Transmission electron microscopy images were taken by FEI (TECNAI) instrument.

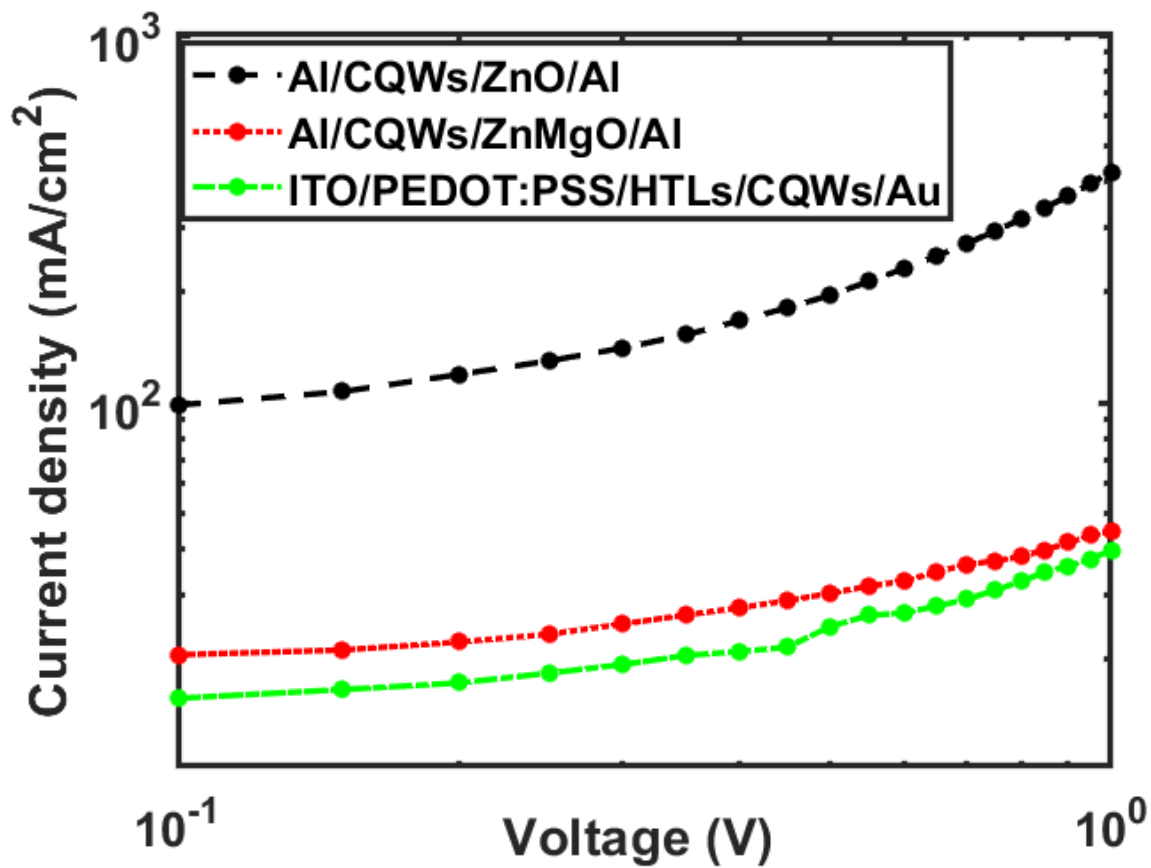

Fig. S1. Current density vs. voltages for the electron-only devices using the structure of Al/CQWs/ZnO/Al and Al/CQWs/ZnMgO/Al and the hole-only device using the structure of ITO/PEDOT:PSS/HTLs/CQWs/Au. Note that the thicknesses of all layers have been kept the same as those of the active device.

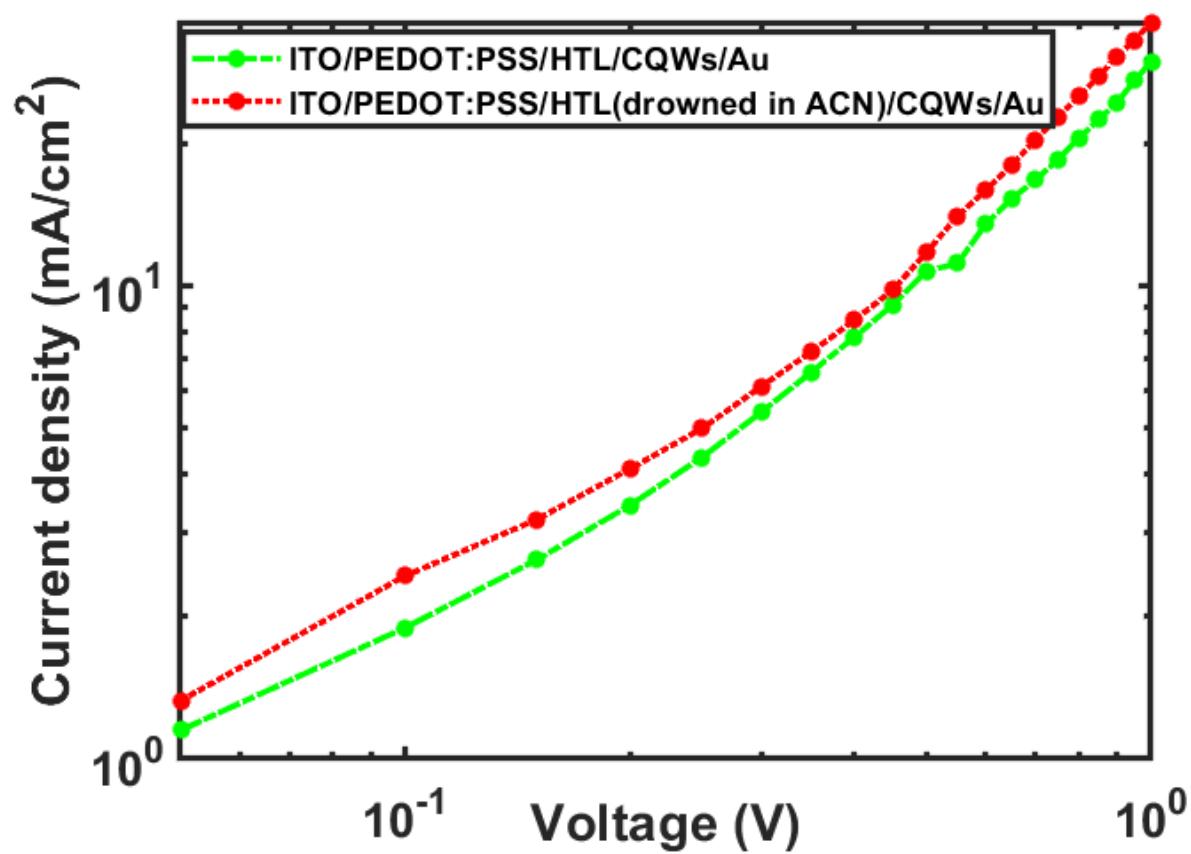

Fig. S2. Current density vs. voltages for the hole-only device in the structure of ITO/PEDOT:PSS/HTLs/CQWs/Au with and without drowning into acetonitrile (ACN).

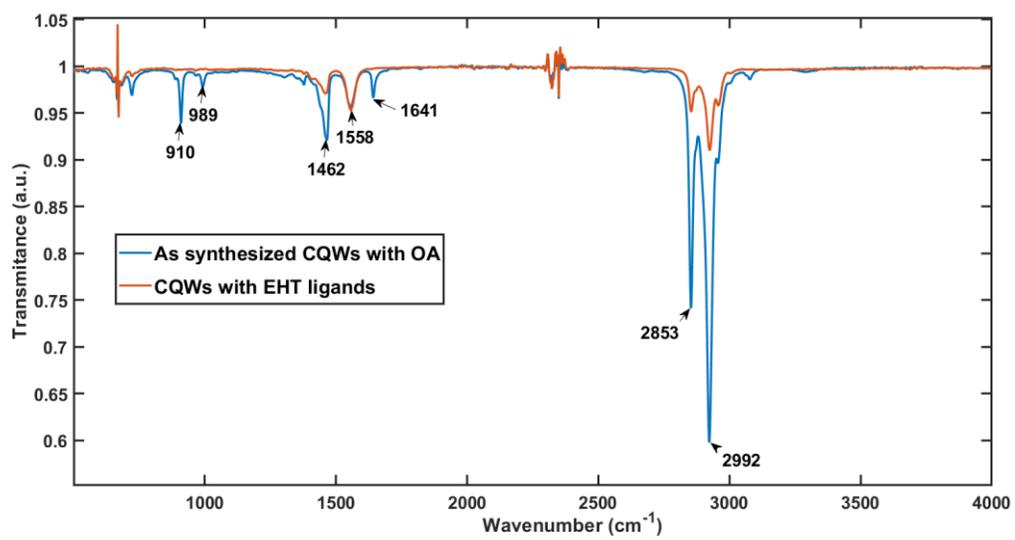

Fig. S3. Fourier-transform infrared spectroscopy (FTIR) spectra of both HIS-CQWs and LE-HIS-CQWs.

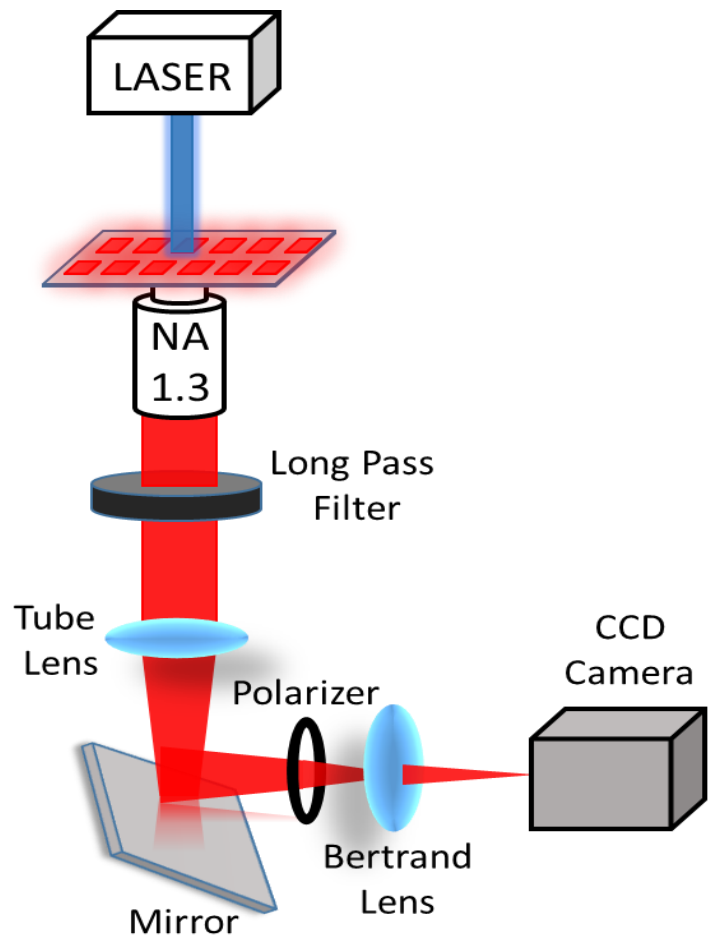

Fig S4. Schematic illustration of back focal plane (BFP) imaging.

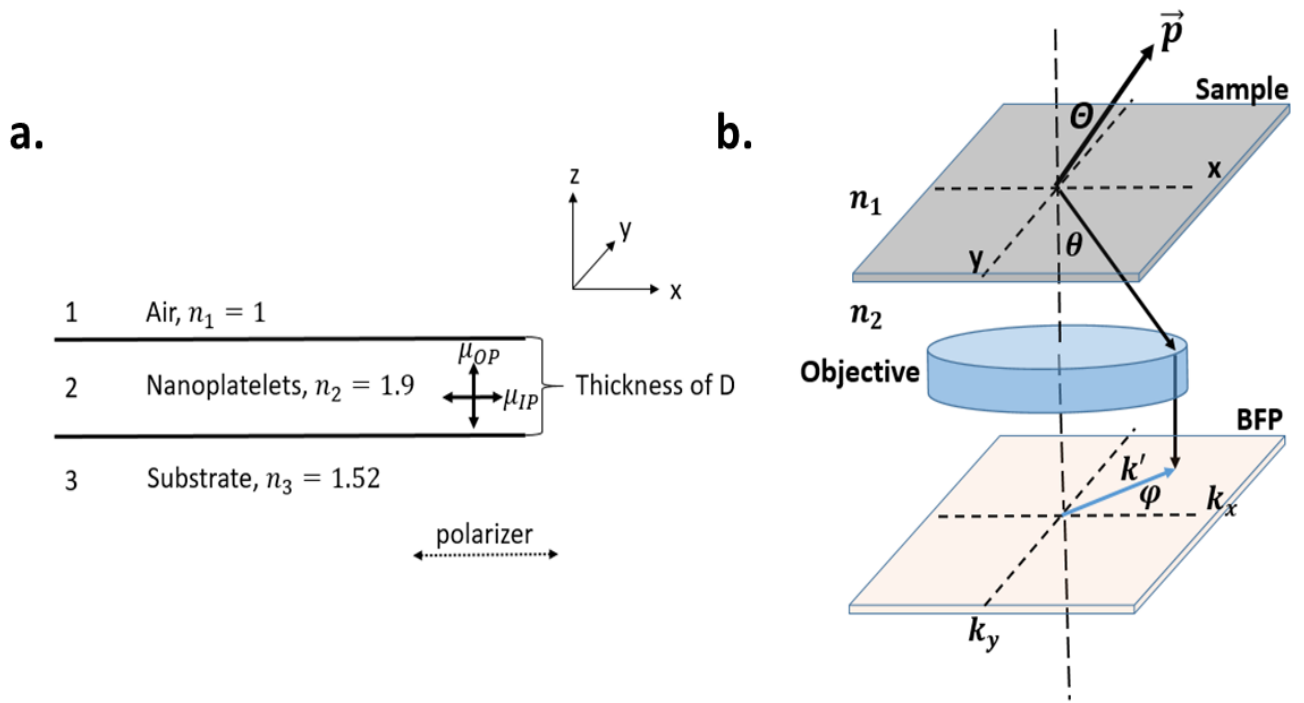

Fig. S5. Schematic of a) the three-layer structure of air, an NPL emitter layer, and a substrate, and b) Emission configuration indicating the coordinate system used for calculations.

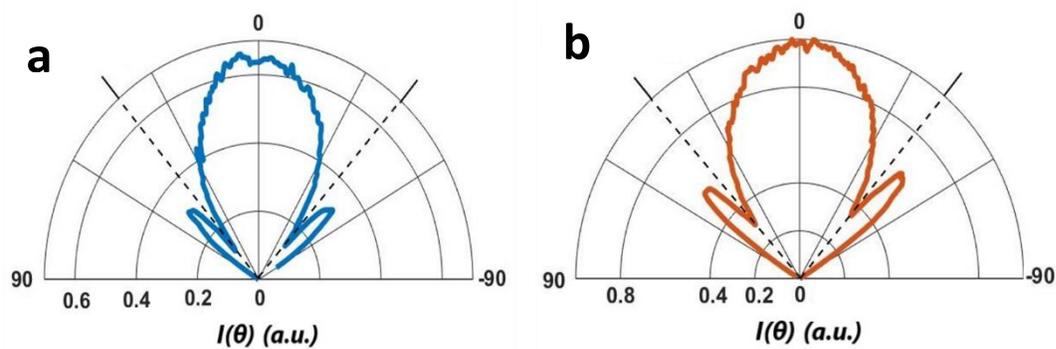

Fig. S6. Angle-dependent intensity profile for a) the self-assembled monolayer (SAM) and b) the spin-casted NPLs.

**Back focal plane (BFP) imaging**

The setup we used is schematically illustrated in fig. S4. Self-assembled and spin-casted film of NPLs on quartz substrates were excited by a 400 nm laser (see fig S5 a-b). Using an inverted optical microscope (Nikon eclipse Ti-U) with a high numerical aperture (NA) objective lens of 1.3 and immersion oil and using a charge-coupled device (CCD) camera (Thorlabs), the angular distribution of fluorescence intensity was collected.

In the BFP pattern, each point corresponds to the elevation angle of emission Θ of the nanoplatelets defined by the photon momentum k (see fig. S5). Theoretically[2], [3], the relation between k and Θ is determined by the equation k=n sin(θ), where n is the refractive index of immersion oil (n=1.52 in our experiment). In addition, the maximum k determines by the NA of the objective, which is k = 1.3. Experimentally, the transition dipole moments (TDMs) direction (Θ) is defined by the ratio of the horizontal ($p_\parallel$) component to the sum of the horizontal and vertical ($p_\perp$) elements (Θ = $p_\parallel$/($p_\parallel$+ $p_\perp$)). Figs. 2a-c show the k-dependent intensity profile associated with the p-polarized direction of dipoles with respect to the polarizer on both theoretically and experimentally. It has been reported that the in-plane (horizontal) orientation dipoles have a distinctive feature: the intensity is zero at $k_x/k_0=1$, where $k_x$ and $k_0$ are the photon momenta parallel to the substrate and in air, respectively. Using this feature, we determined the ratio of horizontal TDMs with respect to the total TDMs.

**Theoretical derivations**

First, the emission pattern and dipole distribution were calculated theoretically using Schuller et al.[3] and Scott et al.[2], which provide detailed information about the derivation of the model. Using their final formulas, we found the ratio of the in-plane (IP) to out-of-plane (OP) TDMs direction. As illustrated in fig. S5a, considering a structure including three layers, air, an emitting layer which is the NPLs, and a substrate as quartz, Fresnel reflection (r) and transmission (t) coefficients describe the plane-wave reflected and transmitted between the interfaces (i, j = 1,2,3):

$$t_{ij}^p = \frac{2n_i n_j k_{zi}}{n_j^2 k_{zi} + n_i^2 k_{zi}}, \quad t_{ij}^s = \frac{2k_{zi}}{k_{zi} + k_{zj}}, \tag{S1}$$

$$r_{ij}^p = \frac{n_j^2 k_{zi} - n_i^2 k_{zi}}{n_j^2 k_{zi} + n_i^2 k_{zi}}, \quad r_{ij}^s = \frac{k_{zi} - k_{zj}}{k_{zi} + k_{zj}}, \tag{S2}$$

where n is the refractive index, k is the perpendicular components of momentum described by the equation $k_{zi} = \sqrt{k_0^2 \epsilon_i - k_x^2}$ where $k_0$ is the wavevector in air, and s and p indicate the polarization of the electromagnetic wave.

Then, the local densities of optical states for IP and OP dipoles were calculated by the effect of reflection on the interfaces and interference inside the emitter layer (figs. S5 a-b),

$$\rho_{IP}^s = p_x^s(k_x, k_y) = \left(\frac{1}{8\pi k_0^2}\right)\left(\frac{k_0}{k_{z3}}\right) \left| \frac{t_{32}^s e^{\frac{ik_{z2}D}{2}}\left(1 + r_{21}^s e^{\frac{2ik_{z2}D}{2}}\right)}{1 - r_{21}^s r_{23}^s e^{2ik_{z2}D}} \frac{k_y}{\sqrt{k_x^2 + k_y^2}} \right|^2 \tag{S3}$$

$$\rho_{IP}^p = p_x^p(k_x, k_y) = \left(\frac{1}{8\pi k_0^2}\right)\left(\frac{k_0}{k_{z3}}\right)\left|\frac{t_{32}^p e^{\frac{ik_{z2}D}{2}} \frac{k_{z2}}{n_2 k_0}\left(1-r_{21}^p e^{\frac{2ik_{z2}D}{2}}\right)}{1-r_{21}^p r_{23}^p e^{2ik_{z2}D}} \frac{k_x}{\sqrt{k_x^2 + k_y^2}}\right|^2 \quad (S4)$$

$$\rho_{OP}^p = p_z^s(k_x, k_y) = \left(\frac{1}{8\pi k_0^2}\right)\left(\frac{k_0}{k_{z3}}\right)\left|\frac{t_{32}^p e^{ik_{z2}D/2} \frac{k_x}{n_2 k_0}\left(1 + r_{21}^p e^{2ik_{z2}D/2}\right)}{1-r_{21}^p r_{23}^p e^{2ik_{z2}D}}\right|^2 \quad (S5)$$

Based on the above equations, the intensity of emission on the back focal plane with respect to IP or OP dipoles population and dipole moment (μ) were calculated as

$$N^s(k_x, k_y) = A\rho_{IP}^s f_{IP}|\mu_{IP}|^2 \quad (S6)$$

$$N^p(k_x, k_y) = A\left(\rho_{IP}^p f_{IP}|\mu_{IP}|^2 + \rho_{OP}^p f_{OP}|\mu_{OP}|^2\right) \quad (S7)$$

where N is the calculated intensity of emission and A is the experimental fitting factor.

Using the above equations, the emission pattern is calculated for %100 IP and %100 OP, as shown in figs. 2 a-b, respectively.

Thus, the difference between 100% IP TDMs to the fitted curve has been measured by fitting the simulation curves to the experiment (see fig. 3b). Here, this difference resulted in the ratio of IP TDMs. The contributions of IP TDMs obtained from theoretically fitting the k-dependent intensity profile of samples are 92% for the self-assembly; it is about 52% deeper than for the spin-coated nanoplatelets (fig. S8 and 3b). In addition, the angle-dependent intensity profile in fig. S6 indicates a more directed emission for self-assembly nanoplatelets by virtue of sharper shape between the oval line and two lobes (see figs. 2 c,d).

**Light extraction efficiency calculations**

The LED structure's light-extraction efficiency (LEE) calculations were carried out using the finite-discrete time-domain (FDTD) method. Simulations were conducted by utilizing commercially available Lumerical FDTD Solutions (ANSYS Inc). To simulate the structure properly, thicknesses and refractive indices of each layer were determined from ellipsometric measurements.

The active region of the LED, which consists of NPLs, was modeled as an electric dipole. The dipole was placed in the center of the active layer. Metal boundary condition was used on the verge of the cathode layer. The other boundary conditions were selected as a perfectly matched layer (PML), which are perfect light absorbers. Since the area of the simulation region must be kept large enough so that fields radiated in the structure do not reach the absorbing boundaries, FDTD calculation was performed in a region with the dimensions of $10\times10\times2.29$ μm$^3$.

The far-field projection was used to obtain the angular distribution of light in the substrate. Transmission coefficients at the interface between substrate and air were calculated using Fresnel's equations[4].

$$t_\perp = \frac{2\sin\theta_t \cos\theta_i}{\sin(\theta_i + \theta_t)} \quad (S8)$$

$$t_\parallel = \frac{2\sin\theta_t \cos\theta_i}{\sin(\theta_i + \theta_t)\cos(\theta_i - \theta_t)} \quad (S9)$$

where $t_\perp$ and $t_\parallel$ are the perpendicular and parallel components of the transmission coefficient, respectively. $\theta_i$ and $\theta_t$ are incident and transmission angles. From incident electric field and transmission coefficients, the angular distribution of the light in the air was found. Extracted power was determined by:

$$P_{ext} = \frac{1}{2}\sqrt{\frac{\epsilon_0}{\mu_0}} \iint |E_{air}(\theta, \phi)|^2 \sin(\theta)\, d\theta d\phi \quad (S10)$$

where $E_{air}(\theta, \phi)$ is the electric field in the air with a specific direction angle.

Total extracted power from the orientation-controlled and that from randomly-oriented structures were calculated by repeating the above analysis for three fundamental dipole directions: ' x',' y', and 'z'. From fig. S7, it can be seen that 'x' and 'y' orientations are parallel to the structure while 'z' orientation is perpendicular. To calculate light extraction efficiency, the results of each of these three analyses were multiplied with the ratio of dipoles in a specific orientation to all dipoles, added to each other, and divided by average dipole power[5].

To determine the ratio of parallel and perpendicular orientations of dipoles in a device, back focal plane (BFP) spectroscopy was used[2]. The results show that while 90% of dipoles in the orientation-controlled structure are parallel to the device, in randomly oriented structure rate of dipoles with parallel orientation is 67%.

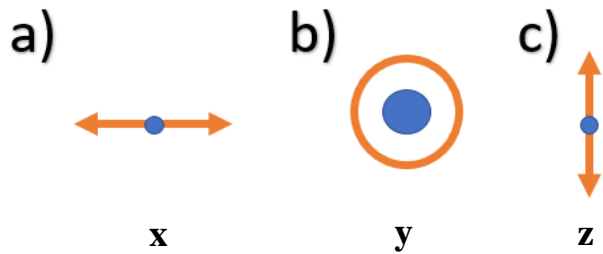

a) x   b) y   c) z

Fig. S7. Schematics of the dipole in a) x-direction, b) y-direction, and c) z-direction. x and y are parallel to the structure, and z is perpendicular.

**References**

[1] G. H. V. Bertrand, A. Polovitsyn, S. Christodoulou, A. H. Khan, and I. Moreels, "Shape control of zincblende CdSe nanoplatelets," *Chem. Commun.*, vol. 52, no. 80, pp. 11975–11978, Sep. 2016.

[2] R. Scott, J. Heckmann, A. V. Prudnikau, A. Antanovich, A. Mikhailov, N. Owschimikow, M. Artemyev, J. I. Climente, U. Woggon, N. B. Grosse, and A. W. Achtstein, "Directed emission of CdSe nanoplatelets originating from strongly anisotropic 2D electronic structure," *Nat. Nanotechnol.* **2017**, vol. 12, no. 12, pp. 1155–1160.

[3] J. A. Schuller, S. Karaveli, T. Schiros, K. He, S. Yang, I. Kymissis, J. Shan, and R. Zia, "Orientation of luminescent excitons in layered nanomaterials," *Nat. Nanotechnol.* **2013**, vol. 8, no. 4, pp. 271–276

[4] E. Hecht, *Optics,* Pearson Education, 5$^{th}$ edition, **2019**.

[5] A. Chutinan, K. Ishihara, T. Asano, M. Fujita, and S. Noda, "Theoretical analysis on light-extraction efficiency of organic light-emitting diodes using FDTD and mode-expansion methods," *Org. Electron.* **2005**, vol. 6, no. 1, pp. 3–9.